\documentstyle[12pt,epsfig]{aipproc}
\begin{document}
\thispagestyle{empty}
\begin{flushright}
\large
DO-TH 97/10 \\ May 1997
\end{flushright}
\vspace{1.2cm}

\begin{center}
\Large
\hbox to\textwidth{\hss
{\bf Next-to-Leading Order Evolution} \hss}

\vspace{0.1cm}
\hbox to\textwidth{\hss
{\bf of Polarized Fragmentation Functions} \hss}

\vspace{1.2cm}
\large
M.\ Stratmann\\
\vspace{0.5cm}
\normalsize
Universit\"{a}t Dortmund, Institut f\"{u}r Physik, \\
\vspace{0.1cm}
D-44221 Dortmund, Germany \\
\vspace{2.6cm}
{\bf Abstract} 
\end{center}

\noindent
We present a brief description of the determination of the two-loop
spin-dependent time-like splitting functions relevant for the
NLO evolution of polarized fragmentation functions. Our calculation
based on the analytic continuation of the corresponding
space-like results obtained within the light-cone gauge method proposed
by Curci, Furmanski, and Petronzio. As an application we present 
an analysis of polarized $\Lambda$  production in $e^+e^-$ and $ep$
collisions. 

\vspace{6.0cm}
\noindent
{\it Talk presented at the  5th International Workshop on
'Deep Inelastic Scattering and QCD' (DIS '97), Chicago, IL, April 14-18, 1997.}
\vfill

\setcounter{page}{0}
\newpage
\title{NLO Evolution of Polarized Fragmentation Functions}
\author{Marco Stratmann}
\vspace*{-0.3cm}
\address{Universit\"at Dortmund, Institut f\"ur Physik\\
D-44221 Dortmund, Germany}
\maketitle
\vspace*{-0.8cm}
\begin{abstract}
We present a brief description of the determination of the two-loop
spin-dependent time-like splitting functions relevant for the
NLO evolution of polarized fragmentation functions. Our calculation
based on the analytic continuation of the corresponding
space-like results obtained within the light-cone gauge method proposed
by Curci, Furmanski, and Petronzio. As an application we present 
an analysis of polarized $\Lambda$  production in $e^+e^-$ and $ep$
collisions. 
\end{abstract}
%
%
\vspace*{-0.3cm}
\section*{General Framework}
%
%
\vspace*{-0.15cm}
\noindent
In analogy to the familiar space-like (S) structure function
$g_1^{(S)}(x,Q^2)$ appearing in longitudinally polarized DIS 
one can define a similar time-like (T) 
structure function $g_1^{(T)} (z,Q^2)$, where $q^2\equiv Q^2>0$,
which describes the helicity transfer in single-particle inclusive 
$e^+ e^-$ annihilation processes (SIA), e.g.,
\begin{equation} 
\frac{d\Delta \sigma^{e^+e^-\rightarrow HX}}{dy dz}\equiv
\frac{d \sigma^{e^{-}(+),H(+)}}{dy dz} -
\frac{d \sigma^{e^{-}(+),H(-)}}{dy dz} =
\frac{6 \pi \alpha^2}{Q^2} (2y-1) g_1^{(T)} (z,Q^2) \; .
\end{equation}
where $z=2 p_H\cdot q/Q^2$, $y=p_H\cdot p_{e^-}/p_H\cdot q$ and
$(\pm)$ denotes the helicities of the $e^-$ and $H$.
In terms of the spin-dependent fragmentation functions 
$\Delta D_f^H (z,Q^2) \equiv D_{f(+)}^{H(+)}(z,Q^2) - D_{f(+)}^{H(-)}(z,Q^2)$, 
where $D_{f(+)}^{H(+)}(z,Q^2)$ ($D_{f(+)}^{H(-)}(z,Q^2)$) is the 
probability for finding a hadron $H$ with positive (negative) helicity in a 
parton $f$ with positive helicity at a mass scale $Q$, carrying a 
fraction $z$ of the parent parton's momentum, we can write 
$g_1^{(T)}$ to NLO as \cite{ref1} 
(the symbol $\otimes$ denotes the usual convolution)
\begin{equation}   
\!\!g_1^{(T)}(z,Q^2)\! = \! \sum_{q} e_q^2 
\Bigg\{ \!\! \left[ \Delta D_q^H + \Delta D_{\bar{q}}^H \right]\! \otimes \! 
\Delta {\cal C}_q^{(T)} \! + 2 \Delta D_g^H 
\! \otimes \! \Delta {\cal C}_g^{(T)} \! \Bigg\} (z,Q^2)\;.
\end{equation}

The $Q^2$-evolution of the $\Delta D_f^H$, predicted by QCD, is 
again similar to the space-like case. 
The singlet evolution equation, e.g., schematically reads
\begin{eqnarray} 
\frac{d}{d\ln Q^2} \left({\Delta D_{\Sigma}^H} \atop {\Delta D_g^H}\right) 
(z,Q^2) = \left[ 
\Delta \hat{P}^{(T)} \otimes 
 \left({\Delta D_{\Sigma}^H} \atop {\Delta D_g^H}\right) \right](z,Q^2) \;,
\end{eqnarray}
with $\Delta D_{\Sigma}^H\equiv \sum_q (\Delta D_q^H + \Delta D_{\bar{q}}^H)$. 
It should be recalled that the off-diagonal entries in
the singlet evolution matrices $\Delta \hat{P}^{(S,T)}$ 
interchange their role when going from the space- to the time-like case,
see, e.g., \cite{ref5a,ref1}.

As a manifestation of the so-called Gribov-Lipatov relation \cite{ref2} the
space- and time-like splitting functions are equal in LO. Furthermore they are 
related by analytic continuation of the space-like splitting functions 
(Drell-Levy-Yan relation (ACR) \cite{ref3}) which can be schematically 
expressed as $(z<1)$
\begin{equation}
\Delta P_{ij}^{(T)}(z) = z {\cal AC} \Bigg[ \Delta P_{ji}^{(S)} 
(x=\frac{1}{z}) \Bigg] \; ,
\end{equation}
where the operation ${\cal AC}$ analytically continues any function to 
$x \rightarrow 1/z >1$ and correctly adjusts the color factor and the sign 
depending on the splitting function 
under consideration \cite{ref1}, e.g.,
\begin{equation} 
\Delta P_{qq}^{(T)}(z) = -z \Delta P_{qq}^{(S)} (\frac{1}{z}) \; ,\;
\Delta P_{gq}^{(T)}(z) = \frac{C_F}{2 T_f} z \Delta P_{qg}^{(S)} 
(\frac{1}{z}) \;,\;
\ldots
\end{equation}
These LO relations are based on symmetries of tree diagrams under 
crossing, and should be therefore no longer valid when going to NLO.
Fortunately \cite{ref4}, the breakdown of the ACR arising beyond LO is 
essentially due to kinematics and can be rather straightforwardly detected 
within the light-cone gauge method used 
in \cite{ref5a,ref4,ref5b,ref6} to calculate the space-like splitting functions. 

%
\vspace*{-0.3cm}
\section*{Understanding of the ACR breaking}
%
\vspace*{-0.15cm}
\noindent
The general strategy of the light-cone gauge method is based on a rearrangement
of the perturbative expansion of a partonic cross section into a finite hard
part and a process-{\em{in}}dependent part $\Gamma_{ij}$ which contains
all (and only) mass singularities \cite{ref7,ref4}. Using dimensional regularization
the $\overline{\rm{MS}}$ 
evolution kernels appear order by order as the residues 
of the {\em{single}} $1/\epsilon$ poles in $\Gamma_{ij}$ \cite{ref4}.
The difference between the space- and the time-like 
$\Gamma_{ij}$-kernels essentially amounts to 
relative extra phase space factors ($z^{-2\epsilon}$) in the time-like
case\footnote{In the unpolarized case
a further difference arises {\protect{\cite{ref1}}} 
through the necessary adjustment of
the different spin-average factors for gluons and quarks in  
$d=4-2\epsilon$ dimensions.} \cite{ref4,ref1}. 
All this gives on aggregate for $z<1$ for the AC
of the spin-dependent space-like kernels \cite{ref1}
\begin{equation} \label{eq1}
\!\!
\Delta \Gamma_{qq}^{(T)}(z,\frac{1}{\epsilon}) \!=\! -z^{1-2 \epsilon} 
\Delta \Gamma_{qq}^{(S)} (\frac{1}{z},\frac{1}{\epsilon}), \;
\Delta \Gamma_{gq}^{(T)}(z,\frac{1}{\epsilon})\!=\! \frac{C_F}{2 T_f} 
z^{1-2 \epsilon} \Delta \Gamma_{qg}^{(S)}
(\frac{1}{z},\frac{1}{\epsilon}),\ldots
\end{equation}
Obviously, higher $(1/\epsilon^2)$ pole terms in $\Delta \Gamma_{ij}^{(S)}$ 
will generate additional contributions to the relevant {\em{single}} pole part
of $\Delta \Gamma_{ji}^{(T)}$ if combined with the factors $z^{-2 \epsilon}$ 
in (\ref{eq1}).
To extract the contributions that break the ACR 
one can easily go through the NLO calculation of the 
$\Delta \Gamma_{ij}^{(S)}$ \cite{ref6} graph by graph picking up all 
$1/\epsilon^2$ pole terms. 
Apart from the slightly subtle case of the so-called 'ladder-subtraction'
topology \cite{ref6} this is a rather straightforward task \cite{ref1}.
To obtain the desired final result for $\Delta P_{ij}^{(T)}$ \cite{ref1} one has
to combine the results of the AC of the NLO splitting functions 
$\Delta P_{ij}^{(S)}$ \cite{ref8,ref6}, by using the operation ${\cal AC} [...]$ 
defined above, with the contributions that break the ACR. 
Finally, the missing endpoint contributions, i.e., the terms $\sim \delta (1-z)$, 
are exactly the same as in the space-like situation \cite{ref4,ref5a}.
In the same way the hard subprocess cross sections $\Delta C_i^{(T)}$ in (2)
can be obtained via analytic continuation (see ref.\ \cite{ref1} for details).
The final results for the $\Delta P_{ij}^{(T)}$ and $\Delta C_i^{(T)}$ are too
long to be listed here but can be found in ref.\ \cite{ref1}\footnote{We also 
have recalculated the unpolarized NLO time-like
Altarelli-Parisi kernels where we fully agree with the results given in
\cite{ref4,ref5a}.}.

The rather simple and transparent structure of the ACR breaking part 
is a hint that there could be a more straightforward 
way of linking the time- and the analytically continued space-like 
NLO quantities.
The key observation for such considerations is to
notice that although the usual 4-dim.\ LO splitting functions obey the ACR rule,
the rule must break down for their $d=(4-2\epsilon)$-dimensional counterparts as 
an immediate consequence of eq.(\ref{eq1}) \cite{ref1}. 
It turns out \cite{ref1} that the breakdown of the ACR beyond LO in 
the $\overline{\rm{MS}}$ scheme is entirely driven by this
breaking such that it can be accounted for by a simple
factorization scheme transformation \cite{ref1}.

Finally, our results for $\Delta P_{ij}^{(T)}$
(as well as the corresponding unpol.\ results given in \cite{ref4,ref5a})
fulfil the so-called SUSY relation which links all 
singlet splitting functions in a remarkably simple way in the limit $C_F=N_C=2 T_f
\equiv N$ (cf.\ \cite{ref8,ref6}) if they are properly transformed to a
regularization method which respects SUSY such as dimensional reduction \cite{ref10}.
The validity of the SUSY relation \cite{ref1} serves as an important
check for the correctness of our results.

%
\vspace*{-0.3cm}
\section*{Application: Polarized $\Lambda$ production}
%
\vspace*{-0.15cm}
\noindent
The most likely candidate for a measurement of 
polarized fragmentation functions is the $\Lambda$ baryon due to its
self-analyzing decay $\Lambda \rightarrow p\pi^-$.
In \cite{ref11} a strategy was proposed for extracting the 
$\Delta D_f^{\Lambda}$ in SIA $e^+ e^- \rightarrow \Lambda X$. 
At high enough energies, no beam polarization is needed 
since the parity-violating coupling $q\bar{q}Z$ automatically generates a
net polarization of the quarks. 
In Fig.\ 1a we compare the first results for such a measurement of the
cross section asymmetry $A^{\Lambda}_{SIA}$ on the
$Z$-pole by ALEPH and DELPHI \cite{ref12} with two
conceivable NLO models for the $\Delta D_f^{\Lambda}$ (the corresponding 
LO results are very similar) \cite{ref15}. 
Scenario 1, where $\Delta D_s^{\Lambda}(z,\mu^2)=z^{\alpha}D_s^{\Lambda}(z,\mu^2)$,
$\Delta D_{f\neq s}^{\Lambda}(z,\mu^2)=0$, is based on the naive quark model
whereas for scenario 2 we also allow for a 
non-vanishing $\Delta D_{u,d}^{\Lambda}$ input as discussed in \cite{ref11}.
In both cases the inputs for the evolution are specified
at some low starting scale $\mu^2\simeq {\cal{O}}(0.3\,\mathrm{GeV}^2)$ 
(for more details see ref.\ \cite{ref15} where also a
model for the unpolarized $D_f^{\Lambda}$ can be found).

Apart from SIA the possibility of extracting the $\Delta D_f^{\Lambda}$ 
in semi-inclusive DIS (SIDIS), i.e.,
$ep \rightarrow e \Lambda X$, has also been discussed recently 
\cite{ref13}. Here, either a longitudinally polarized lepton beam or a 
polarized nucleon target would be required. 
Such kind of measurement appears possible e.g. for the HERMES 
experiment \cite{ref14}. 
In Fig.\ 1b we present LO and NLO predictions for the cross section asymmetry
$A^{\Lambda}_{SIDIS}$ in $\vec{e}p\rightarrow \vec{\Lambda} X$ as a function of $x$ 
at $Q^2=3\,\mathrm{GeV}^2$ and where we have integrated over $0.3\leq z\leq 0.9$ 
\cite{ref15}.
Such a measurement by HERMES would allow, in principle, to discriminate between 
the two proposed scenarios for the polarized $\Lambda$ fragmentation functions
$\Delta D^{\Lambda}_f$.
\vspace*{-0.2cm}
\begin{figure}[h!] 
\centerline{\epsfig{file=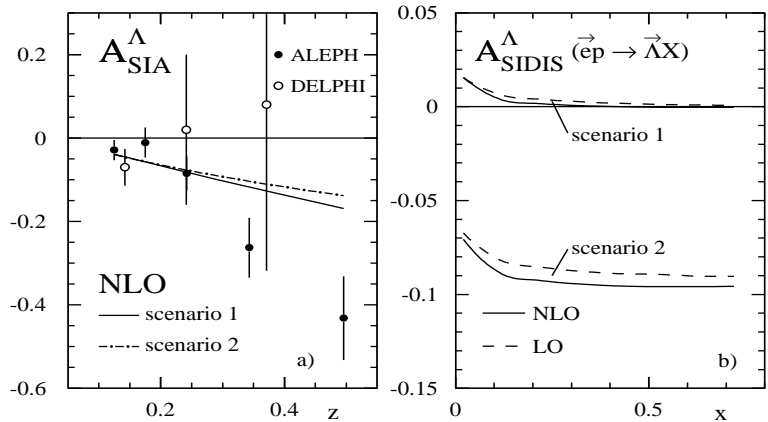,width=11cm,height=6.5cm}}
\vspace*{-0.6cm}
\caption{a) Comparison of recent data for $A^{\Lambda}_{SIA}$ [13] with the 
results obtained for two conceivable scenarios for the $\Delta D_f^{\Lambda}$; 
b) LO and NLO 
predictions for $A^{\Lambda}_{SIDIS}$ in $\vec{e}p\rightarrow \vec{\Lambda} X$.}
\label{fig1}
\end{figure}

\vspace*{-0.3cm}
{\small \noindent {\em{Acknowledgements:}} It is a pleasure to
thank W.\ Vogelsang and D.\ de Florian for a fruitful collaboration. 
This work was partly supported by the BMBF Bonn.}


\end{document}